# Joint Secured and Robust Technique for OFDM Systems

[1,2]A. Al-Dweik, [1]M. Mirahmadi, [1]A. Shami, [2]Z. Ding and [3]R. Hamila
[1]Western University, Canada, e-mail: {aaldweik,mmirahma, ashami2}@uwo.ca
[2]Newcastle University, UK, e-mail: z.ding@newcastle.ac.uk
[3]Qatar University, Qatar, e-mail: hamila@qu.edu.qa

*Abstract*—This work presents a novel technique for joint secured and robust transmission of orthogonal frequency division multiplexing (OFDM) based communication systems. The proposed system is implemented by developing a new OFDM symbol structure based on symmetric key cryptography. At the receiver side, data detection becomes infeasible without the knowledge of the secret key. For an intruder who tries to detect the data without the knowledge of the key, the signal will be a noise-like signal. In addition to the system security, theoretical and simulation results demonstrated that the proposed system provides time and frequency diversity, which makes the system highly robust against severe frequency-selective fading as well as other impairments such as impulsive noise and multiple access interference. For particular frequency-selective fading channels, the bit error rate (BER) improvements was about $15$ dB at $BER \sim 10^{-4}$.

*Index Terms*— OFDM, security, encryption, physical layer.

## I. INTRODUCTION

Although OFDM is highly robust against various transmission impairments, it does not have any inherent security features. Hence, additional encryption/decryption algorithms should be implemented to grant data security. Generally speaking, most data security algorithms are implemented in the higher layers of the OSI model. For example, the WEP protocol is integrated as a part of the data-link layer (layer 2), the IPSec is part of the network layer (layer 3), and most widely used block and stream ciphers are implemented in layers 5, 6 and 7 [1], [2]. The main limitation of such commonly designed ciphers is their low speed, which is a major drawback for today's broadband systems. Consequently, substantial research efforts were recently steered towards physical-layer security (PLS) algorithms, several examples for PLS systems can be found in [3], [4] and the references listed therein.

Generally speaking, most PLS-OFDM cryptosystems are designed to function at the bit or symbol level. For example, Dzung [5] developed a system which encrypts the baseband QAM symbols by changing their phase according to a given key sequence before the Inverse Fast Fourier transform (IFFT) process. Moreover, the training symbols that are embedded for synchronization and channel estimation are encrypted as well. Consequently, the encryption process hides the necessary information required for synchronization and channel estimation, which are necessary to recover the encrypted data symbols. Furthermore, the data symbols themselves are encrypted as well. Moffatt [6]-[7] developed a security system for OFDM by mixing the phases of the data symbols and varying the data-to-subcarrier assignment based on a secret key sequence. Therefore, an eavesdropper needs first to know the mapping between data and subcarriers, and then the phase/amplitude of the data symbols. Similar to the above mentioned systems, there are several other systems based on the general concept of building an encryption technique by processing the frequency domain symbols [8], [9]-[11].

In this work we exploit the OFDM unique structure to develop a secured communication system that relies on the sensitivity of OFDM systems to synchronization errors [12], [13]. At the transmitter side, the encryption is performed by hiding some necessary synchronization information based on a secrete key sequence. At the receiver side, the decryption would be impossible without the knowledge of the correct synchronization parameters. For an attacker who does not have the correct key, the received signal will be a noise-like signal. Unlike most PLS-OFDM systems, the proposed system utilizes the time-domain samples post the IFFT. Consequently, the new system will have strong security as well as low computational complexity. Furthermore, the proposed system has high immunity against multipath fading, jamming and impulsive noise due to the time and frequency diversity inherent to this system.

In what follows unless otherwise specified, uppercase boldface and blackboard letters such as $\mathbf{H}$ and $\mathbb{H}$ will denote $N \times N$ matrices, whereas lowercase boldface letters such as $\mathbf{x}$ will denote row or column vectors with $N$ elements. Bold upper and lowercase letters with a tilde such as $\tilde{\mathbf{H}}$ and $\tilde{\mathbf{r}}$ will denote $N_t \times N_t$ matrices and row/column vectors with $N_t$ elements, respectively. Symbols with a hat such as $\hat{x}$ will denote the estimate of $x$. Encrypted vectors will be enclosed in angle brackets such as $\langle \mathbf{x} \rangle$.

## II. OFDM SYSTEM DESCRIPTION

In OFDM, the data sequence $\mathbf{d} = [d_0, d_1, ..., d_{N-1}]^T$ is used to modulate $N$ orthogonal subcarriers during the $l$th OFDM symbol, $l = 0, ..., L-1$. In general, the elements of $\mathbf{d}$ are complex symbols drawn uniformly from a quadrature amplitude modulation (QAM) constellations. The modulation process can be implemented efficiently by applying $\mathbf{d}$ to an $N$-points IFFT process to produce the time-domain sequence $\mathbf{x}$, $\mathbf{x} = [x_0, x_1, ..., x_{N-1}]^T$,

$$\mathbf{x}(l) = \mathbf{F}^H \mathbf{d}(l) \quad (1)$$

where the IFFT matrix $\mathbf{F}^H$ is the Hermitian transpose of the normalized $N \times N$ FFT matrix. The elements of $\mathbf{F}^H$ can be expressed as $F_{n,i}^H = \mathcal{E}^{ni}/\sqrt{N}$, $\mathcal{E} \triangleq \exp(j2\pi/N)$, $\{n,k\} \in 0,1,...,N-1$. Therefore, the $n$th sample of $\mathbf{x}(l)$ can be expressed as,

$$x_n(l) = \frac{1}{\sqrt{N}} \sum_{i=0}^{N-1} d_i(l) \mathcal{E}^{ni}, \quad n = 0, 1, ... N-1. \quad (2)$$

Assuming that the CP consists of $N_{CP}$ samples, the transmitted $l$th OFDM block will be composed of $N_t = N + N_{CP}$ samples with the following frame structure,

$$\tilde{\mathbf{x}} = [x_{N-N_{CP}},..., x_{N-1}, x_0, ..., x_{N-1}]^T. \quad (3)$$

After dropping the $N_{CP}$ samples, the received sequence $\mathbf{y} = [y_0, y_1,...,y_{N-1}]^T$ is obtained,

$$\mathbf{y}(l) = \mathbb{H}(l)\, \mathbf{x}(l) + \mathbf{z}(l), \quad (4)$$

The system noise $\mathbf{z} = [z_0, z_1, ..., z_{N-1}]^T$ is modeled as a white Gaussian process with zero mean and variance $\sigma_z^2 = E[|z_n|^2]$, $\mathbb{H}$ denotes the channel matrix during the $l$th OFDM block. Assuming that the channel remains fixed for one OFDM symbol period, the matrix $\mathbb{H}$ can be expressed as

$$\mathbb{H} = \begin{bmatrix} h_0 & h_{N-1} & \cdots & h_1 \\ h_1 & h_0 & \ddots & \vdots \\ \vdots & \vdots & \ddots & h_{N-1} \\ h_{N-1} & h_{N-2} & \cdots & h_0 \end{bmatrix}. \quad (5)$$

The discrete-time channel impulse response samples $h_p = 0$ $\forall\, p \geq L_h$ where $L_h$ is the channel order and $L_h < N_{CP}$. Since $\mathbb{H}$ is a circulant matrix, hence it can be diagonalized by the IFFT/FFT matrices, i.e., $\mathbf{F}\mathbb{H}\mathbf{F}^H = \mathbf{H}$ where $\mathbf{H}$ is $N \times N$ diagonal matrix whose $i$th diagonal element can be expressed as,

$$H_i(l) = \sum_{p=0}^{L_h-1} h_p(l) \mathcal{E}^{-p(N-i+1)}. \quad (6)$$

Therefore, the received samples can be written as

$$\mathbf{y}(l) = \mathbf{F}^H \mathbf{H}(l)\, \mathbf{d}(l) + \mathbf{z}(l) \quad (7)$$

where the $n$th element of $\mathbf{y}$ can be expressed as

$$y_n(l) = \frac{1}{\sqrt{N}} \sum_{i=0}^{N-1} d_i(l) H_i(l) \mathcal{E}^{ni} + z_n(l). \quad (8)$$

The $N$ time-domain samples are then applied to the FFT to produce the decision variables that will be used to reproduce the transmitted symbols, $\mathbf{Y}(l) = \mathbf{F}\,\mathbf{y}(l)$, where .

$$Y_k(l) = d_k(l) H_k(l) + \eta_k(l) \quad (9)$$

where $\eta_k = \frac{1}{\sqrt{N}} \sum_{n=0}^{N-1} z_n \mathcal{E}^{-nk}$ is a Gaussian random variable with the same statistical properties as $z_n$.

## III. THE PROPOSED SYSTEM

As it can be noted from (4), each sample $x_n$ is formed by mixing all data symbols $d_0, ..., d_{N-1}$ using the IFFT. To recover the data symbols, all the time-domain samples $\{x_n, n = 0, \cdots, N-1\}$ should be available at the receiver. Hiding the original values of $x_n$ or loosing part of them will destroy the subcarriers orthogonality and hence inter-carrier interference (ICI) and inter-symbol interference (ISI) will be introduced, which might lead to high bit error rate (BER). Consequently, if a particular function $\mathcal{F}$ is used to hide the original values of $x_n$, it will be difficult to recover the data symbols $d_k$ without errors if $\mathcal{F}$ is not inverted at the receiver side. Therefore, if $\mathcal{F}$ is chosen such that it cannot be guessed by the attackers, and the information loss due to the absence of information about $\mathcal{F}$ is maximized, the data sequence $\mathbf{d}$ can be considered to be secured. However, the samples $x_n$ are analog, $-N^{-\frac{1}{2}} \leq |x_n| \leq N^{\frac{1}{2}}$, hence one-way functions based on discrete mathematics, which are used in conventional security systems may not be suitable for this system. Moreover, the function $\mathcal{F}$ should be selected such that the average signal-to-noise ratio (SNR) at the receiver side does not degrade when the inverse function $\mathcal{F}^{-1}$ is applied, i.e., the encryption-decryption processes should be distortionless. Inspired by our previous work on OFDM synchronization [12]-[13], we notice that ICI and ISI cannot be avoided unless the samples $x_n$ $\forall$ $n$ belong to the same OFDM symbol $l$, $n \in \{0, 1, ..., N-1\}$, and all samples are placed in the right order, i.e., $x_0$, $x_1$,..., $x_{N-1}$.

Obviously, the ICI/ISI power is proportional to the number of samples included from other OFDM blocks, and/or the number of out of order samples [13]. For a perfect channel ($\sigma_z^2 = 0$ and $\mathbf{H} = \mathbf{I}$) where the $N$-samples block comprises samples from other OFDM blocks or out of order samples, the FFT output can be expressed as

$$d_k(l) = \alpha_k(l) d_k(l) + \beta_k(l) \quad (10)$$

where $\alpha$ is an attenuation factor and $\beta$ is the interference. Both $\alpha$ and $\beta$ depend on the number of samples that belong to the $l$th OFDM symbol, and how many samples are in correct order. Consequently, the useful data symbols at the output of the FFT will be significantly attenuated and heavily buried in interference. Therefore, we can propose to secure the transmitted data by either reordering the time-domain samples $x_n(l)$, mixing the samples of $L$ different OFDM symbols, or both mixing and reordering the samples of $L$ OFDM symbols.

One simple approach to change the order of $x_n$ samples and to mix them with the samples of other OFDM blocks is to use permutation matrices, which can be realized as a random block or convolutional interleaver. Therefore, we assume that before transmission, an $L \times N$ time-domain samples that corresponds to $L$ OFDM symbols collected at the IFFT output are permuted according to an $LN \times LN$ pseudorandom permutation matrix $\mathbf{P}$. Therefore, the encrypted time-domain samples can be expressed as

$$\langle \mathbf{X}(\ell) \rangle = \mathcal{F}(\mathbf{x}(1), ..., \mathbf{x}(L), \mathbf{P}(\ell)) \quad (11)$$
$$= \mathbf{P}(\ell)[\mathbf{x}(1), ..., \mathbf{x}(L)]^T \quad (12)$$

where $\ell$ is the encrypted block index, $\mathbf{P}$ is a permutation matrix that consists of $LN$ ones and $LN(LN-1)$ zeros, and the $LN$ ones are distributed such that no row or column has more than a single one. Subsequently, the $LN$ samples are fragmented into blocks each of which consists of $N$ samples, and the rest of the transmission process is identical to the standard OFDM. Therefore, the received signal after removing the CP samples is given by

$$\langle \mathbf{y}(l) \rangle = \mathbb{H}(l) \langle \mathbf{x}(l) \rangle + \mathbf{z}(l). \quad (13)$$

It can be noted from (13) that the channel matrix should be estimated and compensated before eliminating the permutation process effect. Consequently, the received samples after compensating the channel matrix can be expressed as,

$$\hat{\mathbf{r}}(l) = \hat{\mathbb{H}}^{-1}(l) \langle \mathbf{y}(l) \rangle + \eta(l) \quad (14)$$

where $\eta(l) = \hat{\mathbb{H}}^{-1}(l)\mathbf{z}(l)$. Although the equalization process seems to require time-domain equalization, frequency domain equalization can be applied as well, as described in Section V. Once the channel effects are compensated for $L$ received OFDM symbols, the vector $R(\ell) = [\hat{\mathbf{r}}(1), ..., \hat{\mathbf{r}}(L)]$ is then de-permuted using $\mathbf{P}^{-1}(\ell)$.

## IV. SECURITY ANALYSIS

Security analysis for such a system can be expressed in terms of the probability of estimating the matrix $\mathbf{P}$, and by measuring the information loss due to the inaccurate knowledge of $\mathbf{P}$. In this work we consider the chosen plain-text attack since it is usually considered as one of the strongest measures for security assessment. This type of attacks is based on the assumption that the attacker can choose arbitrary pairs of plain-text and its corresponding cipher-text. In other words, the attacker can decide what is the value of $\mathbf{d}$, and knows $\tilde{\mathbf{y}}(l)$ perfectly. To simplify the discussion, we consider first the case of perfect channel where $\mathbf{H} = \mathbf{I}$ and $\mathbf{z} = \mathbf{0}$. Moreover, the permutation matrix $\mathbf{P}$ is selected to permute only $N$ samples. the received sequence (13) can be expressed as

$$\langle \mathbf{y}(l) \rangle = \mathbf{P}(l)\mathbf{x}(l) = \mathbf{P}(l)\mathbf{F}^H \mathbf{d}(l) \quad (15)$$

Since it was assumed that the data symbols are known, a possible approach to find $\mathbf{P}$ can be achieved by noting that the $n$th element of the right-hand side of (15) can be expressed as

$$\langle y_n(l) \rangle = \sum_{i=0}^{N-1} p_{n,i}(l) x_i(l), \, n \in 0, \cdots, N-1. \quad (16)$$

However, since $p_{n,i} \in \{0,1\}$, and only one out of the $N$ elements in the $n$th row of $\mathbf{P}$ is not zero. Thus (16) can be simplified to

$$\langle y_n(l) \rangle = p_{n,k}(l) x_k(l), \; k \in 0, \cdots, N-1. \quad (17)$$

Hence $p_{n,k}(l) = 1$ and $p_{n,i}(l) = 0 \; \forall \; i \neq k$.

However, at the physical layer there are various impairments that affect the cipher-text $\langle \mathbf{y}(l) \rangle$ such as the AWGN. In such a case, the $n$th sample of the cipher-text can be written as

$$\langle y_n(l) \rangle = \sum_{i=0}^{N-1} p_{n,i}(l) x_i(l) + z_n. \quad (18)$$

This case is more involved due to the noise impact on the cipher-text. However if the matrix $\mathbf{P}$ is fixed for large number of symbols, the attacker can repeat the transmission and reception of $\tilde{\mathbf{y}}$ for $K$ times to eliminate the AWGN,

$$\lim_{l \to K \gg 1} \langle y_n(l) \rangle = p_{n,i}(l) x_i(l) + \frac{1}{K} \sum_{l=0}^{K-1} z_n(l) \approx p_{n,i}(l) x_i(l). \quad (19)$$

Hence, the matrix $\mathbf{P}$ in the presence of noise can be found following the same approach used for the noiseless case, except for the additional averaging. Consequently, a necessary condition for the security of this approach is to make $\mathbf{P}(l) \perp \mathbf{P}(k) \; \forall \; l \neq k, \; [l,k] \in \{1,2,...,L\}$, where $\perp$ denotes the statistical independence. This goal can be achieved by generating $\mathbf{P}$ in a pseudo random fashion based on the encryption key.

Under the assumption that the permutation matrices are independent, the minimum mean-squared error (MMSE) estimation can be used to estimate $\mathbf{P}$,

$$\hat{\mathbf{P}}(l) = \arg \min_{\mathbf{P}_i(l)} |\langle \mathbf{y}(l) \rangle - \mathbf{P}_i(l)\mathbf{x}(l)|^2, \, i = 0, 1, ..., N! - 1 \quad (20)$$

As it can be noted from (20), finding the matrix $\mathbf{P}$ using brute-force approach is quite costly even for the simple $N \times N$ permutation matrix because such process requires $N!$ for the $N \times N$ $\mathbf{P}$ matrix, and $LN!$ trials for the $LN \times LN$ $\mathbf{P}$ matrix. For certain OFDM applications where $N \geq 256$, we can set $L = 1$ because $256! > 2^{1683}$. Thus, it is computationally infeasible to break this system by exhaustive search techniques to find $\mathbf{P}$ even for small size permutation matrices. Another possible approach to enhance the security of the proposed system is to invoke nonlinear $\mathcal{F}$ functions. However, the impact of such functions on the SNR should be investigated carefully.

## V. BER PERFORMANCE IN FADING CHANNELS

Generally speaking, it is infeasible to deduce a general BER solution for all values of $L$ and $\mathbf{P}$. However, extensive simulation results demonstrated that the performance is almost identical for $L \geq 5$ regardless of the matrix $\mathbf{P}$. Moreover, selecting the matrix $\mathbf{P}$ randomly can lead to substantially complicated analysis. Consequently, we choose a special $\mathbf{P}$ matrix to facilitate the analysis, extending the analysis to different $\mathbf{P}$ matrices can be achieved by following the same approach used for this matrix.

Consider the case of $N^2 \times N^2$ $\mathbf{P}$ matrix that is designed so that the first transmitted $N$ samples are the first samples of the $N$ buffered OFDM blocks, the second transmitted $N$ samples are the second samples of the buffered $N$ OFDM blocks, etc. Therefore, the transmitted sequence of $N$ symbols can be expressed as $\langle \mathbf{x}(0) \rangle, \langle \mathbf{x}(1) \rangle,..., \langle \mathbf{x}(N-1) \rangle$ where

$$\langle \mathbf{x}(l) \rangle = \begin{bmatrix} x_l(0) & x_l(1) & \cdots & x_l(N-1) \end{bmatrix}^T. \quad (21)$$

The $N_{CP}$ samples of the CP is formed by copying the last $N_{CP}$ samples of $\langle \mathbf{x} \rangle$ and appending them at the front of $\langle \mathbf{x} \rangle$ to compose the transmission symbol. Consequently, the $i$th transmitted sample is given by

$$\langle \tilde{x}_i(l) \rangle = \langle x_l([i-N_{CP}]_N) \rangle, \quad i = 0, 1, ..., N_t - 1 \quad (22)$$

where $[i]_k \triangleq i \mod k$. The sequence $\langle \tilde{\mathbf{x}} \rangle$ is upsampled, filtered and up-converted to a radio frequency centered at $f_c$ before transmission through the antenna.

At the receiver front-end, the received signal is down-converted to baseband and sampled at a rate $T_s = T_t/N_t$. In this work we assume that the channel is composed of $L_h + 1$ independent multipath components each of which has a gain $h_m$ and delay $m \times T_s$, where $m \in \{0, 1,..., L_h\}$. The channel taps are assumed to be constant over $N$ OFDM symbols, which corresponds to a static multipath channel. The received sequence $\langle \tilde{\mathbf{y}} \rangle$ consists of $N_t$ samples, and it can be expressed as

$$\langle \tilde{\mathbf{y}}(l) \rangle = \tilde{\mathbb{H}}(l) \langle \tilde{\mathbf{x}}(l) \rangle + \tilde{\mathbf{z}}(l), \quad (23)$$

where the channel matrix $\tilde{\mathbb{H}}$ is an $N_t \times N_t$ Toeplitz matrix with $h_0$ on the principal diagonal and $h_1,..., h_{L_h}$ on the minor diagonals, respectively. Given that $L_h + 1 < N_{CP}$, The $n$th sample of $\langle \tilde{\mathbf{y}}(\ell) \rangle$ is given by

$$\langle \tilde{y}_n \rangle = \sum_{i=0}^{L_h - 1} h_i \langle x_{[n-i]_N} \rangle, \quad 0 \leq n < N_t - 1. \quad (24)$$

Subsequently, the receiver should identify and extract the sequence $\langle \tilde{\mathbf{y}} \rangle = \langle [\tilde{y}_{N_{CP}}, \tilde{y}_{N_{CP}+1},..., \tilde{y}_{N+N_{CP}-1}] \rangle$ and discard the CP samples $\langle [\tilde{y}_0, \tilde{y}_1,..., \tilde{y}_{N_{CP}-1}] \rangle$. Therefore,

$$\langle y_n \rangle = \sum_{i=0}^{L_h - 1} h_i \langle x_{[n-i+N_{CP}]_N} \rangle, \quad 0 \leq n \leq N - 1. \quad (25)$$

In vector notation, the sequence $\langle \mathbf{y} \rangle$ can be expressed as,

$$\langle \mathbf{y} \rangle = \mathbb{H} \langle \mathbf{x} \rangle + \mathbf{z}. \quad (26)$$

By noting that $\mathbb{H} = \mathbf{F}^H \mathbf{H} \mathbf{F}$, where

$$\mathbf{H} = \text{diag}\left([H_0, H_1, \cdots, H_{N-1}]\right), \quad H_k = \sum_{m=0}^{L_h} h_m e^{\frac{-j2\pi mk}{N}},$$

then (26) can be written as

$$\langle \mathbf{y} \rangle = \mathbf{F}^H \mathbf{H} \mathbf{F} \langle \mathbf{x} \rangle + \mathbf{z}. \quad (27)$$

To extract $\langle \mathbf{x} \rangle$ from $\langle \mathbf{y} \rangle$ we multiply by the FFT matrix, the inverse of the channel matrix and the IFFT matrix. Therefore, the equalizer output is computed as

$$\langle \mathbf{s} \rangle = \mathbf{F}^H \mathbf{H}^{-1} \mathbf{F} \langle \mathbf{y} \rangle = \langle \mathbf{x} \rangle + \underbrace{\mathbf{F}^H \mathbf{H}^{-1} \mathbf{F}}_{\mathbb{V}} \mathbf{z}. \quad (28)$$

Once again, using the fact that $\mathbb{H} = \mathbf{F}^H \mathbf{H} \mathbf{F}$, we conclude that the matrix $\mathbb{V}$ is circulant where the first row is given by

$$V(0,:) = \frac{1}{N} \begin{bmatrix} \sum_{k=0}^{N-1} \frac{1}{H_k} & \sum_{k=0}^{N-1} \frac{\mathcal{E}^{-k}}{H_k} & \cdots & \sum_{k=0}^{N-1} \frac{\mathcal{E}^{-(N-1)k}}{H_k} \end{bmatrix}. \quad (29)$$

It is interesting to note from (29) that each element in $\mathbb{V}$ is composed of a mixture of all elements of $\mathbf{H}^{-1}$. Consequently, $\mathbb{V}\mathbf{z} \triangleq \mathbf{\Phi}$ can be written as

$$\mathbf{\Phi} = \frac{1}{N} \begin{bmatrix} \sum_{i=0}^{N-1} \sum_{k=0}^{N-1} \frac{z_i}{H_k} \mathcal{E}^{-ik} \\ \sum_{i=0}^{N-1} \sum_{k=0}^{N-1} \frac{z_i}{H_k} \mathcal{E}^{-k[N-1+i]_N} \\ \sum_{i=0}^{N-1} \sum_{k=0}^{N-1} \frac{z_i}{H_k} \mathcal{E}^{-k[N-2+i]_N} \\ \vdots \\ \sum_{i=0}^{N-1} \sum_{k=0}^{N-1} \frac{z_i}{H_k} \mathcal{E}^{-k[i+1]_N} \end{bmatrix}. \quad (30)$$

The decryption process starts by buffering $N$ OFDM symbols, and then multiplying them by $\mathbf{P}^{-1}$. The output depends on the transmitted symbol number $l$ and can be expressed as,

$$\mathbf{s}(l) = \mathbf{x}(l) + \langle \mathbf{\Phi}(l) \rangle = \mathbf{F}^H \mathbf{d}(l) + \langle \mathbf{\Phi}(l) \rangle \quad (31)$$

where

$$\langle \mathbf{\Phi}(l) \rangle = \frac{1}{N} \begin{bmatrix} \sum_{i=0}^{N-1} \sum_{k=0}^{N-1} \frac{z_i(0)}{H_k(0)} \mathcal{E}^{-k[i+N-l]_N} \\ \sum_{i=0}^{N-1} \sum_{k=0}^{N-1} \frac{z_i(1)}{H_k(1)} \mathcal{E}^{-k[i+N-l]_N} \\ \vdots \\ \sum_{i=0}^{N-1} \sum_{k=0}^{N-1} \frac{z_i(N-1)}{H_k(N-1)} \mathcal{E}^{-k[i+N-l]_N} \end{bmatrix}.$$

Applying the FFT to extract the data gives

$$\mathbf{d}^s(l) = \mathbf{F} \left( \mathbf{F}^H \mathbf{d}(l) + \langle \mathbf{\Phi}(l) \rangle \right)$$
$$= \mathbf{d}(\ell) + \mathbf{F} \langle \mathbf{\Phi}(\ell) \rangle \quad (32)$$

where, for example, $\langle \mathbf{\Psi}(0) \rangle \triangleq \mathbf{F} \langle \mathbf{\Phi}(0) \rangle$ can be written as

$$\langle \mathbf{\Psi}(0) \rangle = \frac{1}{N^{\frac{3}{2}}} \begin{bmatrix} \sum_{\mathbf{c}_1=0}^{N-1} \frac{z_i(\ell)}{H_k(\ell)} \mathcal{E}^{-ik}, & \sum_{\mathbf{c}_1=0}^{N-1} \frac{z_i(\ell)}{H_k(\ell)} \mathcal{E}^{-i(k-\ell)}, \\ \cdots, & \sum_{\mathbf{c}_1=0}^{N-1} \frac{z_i(\ell)}{H_k(\ell)} \mathcal{E}^{-i(k-\ell^2)} \end{bmatrix} \quad (33)$$

and where $\sum_{\mathbf{c}_1=0}^{N-1} \triangleq \sum_{i=0}^{N-1} \sum_{\ell=0}^{N-1} \sum_{k=0}^{N-1}$. Hence, while the equalized FFT output of a standard OFDM are given by

$$d_0^s = d_0 + \frac{1}{H_0 \sqrt{N}} \sum_{i=0}^{N-1} z_i \quad (34)$$

for the TDI it is given by,

$$d_0^s = d_0 + \langle \psi_0 \rangle \tag{35}$$

$$= d_0 + \frac{1}{N^{\frac{3}{2}}} \sum_{\ell=0}^{N-1} \sum_{i=0}^{N-1} z_i(\ell) \sum_{k=0}^{N-1} \frac{\mathcal{E}^{-ik}}{H_k(\ell)}. \tag{36}$$

It can be noted from (36) that the decryption process and the last FFT operations have mixed the channel frequency response parameters $H_i$ of the entire interleaving block length. Moreover, since we assumed that the channel is static, then $H_k(l) = H_k \; \forall l$, thus

$$d_0^s = d_0 + \frac{1}{N^{\frac{3}{2}}} \sum_{\ell=0}^{N-1} \sum_{i=0}^{N-1} z_i(l) \sum_{k=0}^{N-1} \frac{\mathcal{E}^{-ik}}{H_k}. \tag{37}$$

The SNR for a given channel matrix $\mathbf{H}$ can be expressed as

$$\text{SNR}|_{\mathbf{H}} = \frac{\sigma_d^2}{\frac{1}{N^3} E\left\{ \left| \sum_{\mathbf{c}_1=0}^{N-1} z_i(\ell) \frac{\mathcal{E}^{-ik}}{H_k} \right|^2 \right\}} \tag{38}$$

where $\sigma_d^2 = E\left\{|d_0|^2\right\}$. The $E\{\cdot\}$ in the denominator of (38) can be expanded to

$$\sum_{\mathbf{c}_2=0}^{N-1} \frac{\mathcal{E}^{-ik}}{H_k} \frac{\mathcal{E}^{\beta\gamma}}{H_\gamma^*} E\left\{ z_i(\ell) z_\beta^*(\alpha) \right\} \tag{39}$$

where $\sum_{\mathbf{c}_2=0}^{N-1} \triangleq \sum_{i=0}^{N-1} \sum_{l=0}^{N-1} \sum_{k=0}^{N-1} \sum_{\alpha=0}^{N-1} \sum_{\beta=0}^{N-1} \sum_{\gamma=0}^{N-1}$. For $l = \alpha$, $i = \beta$ (39) can be reduced to

$$N\sigma_Z^2 \sum_{i=0}^{N-1} \sum_{k=0}^{N-1} \sum_{\gamma=0}^{N-1} \frac{\mathcal{E}^{-i(k-\gamma)}}{H_k H_\gamma^*}. \tag{40}$$

Moreover, since

$$\sum_{i=0}^{N-1} \frac{\mathcal{E}^{-i(k-\gamma)}}{H_k H_\gamma^*} = 0, \text{ for } k \neq \gamma, \tag{41}$$

thus;

$$\sum_{i=0}^{N-1} \sum_{k=0}^{N-1} \sum_{\gamma=0}^{N-1} \frac{\mathcal{E}^{-i(k-\gamma)}}{H_k H_\gamma^*} = 0 \text{ for } k \neq \gamma \tag{42}$$

Therefore,

$$\sum_{\mathbf{c}_2=0}^{N-1} \frac{\mathcal{E}^{-ik}}{H_k} \frac{\mathcal{E}^{\beta\gamma}}{H_\gamma^*} E\left\{ z_i(\ell) z_\beta^*(\alpha) \right\} = \begin{cases} N^2 \sigma_Z^2 \sum_{k=0}^{N-1} \frac{1}{|H_k|^2}, & \begin{matrix} l = \alpha \\ i = \beta \\ k = \gamma \end{matrix} \\ 0, & Otherwise \end{cases}. \tag{43}$$

Finally, the conditional SNR can be expressed as

$$\text{SNR}|_{\mathbf{H}} = \frac{\text{snr}}{\frac{1}{N} \sum_{k=0}^{N-1} \frac{1}{|H_k|^2}} \tag{44}$$

where $\text{snr} = \sigma_d^2/\sigma_Z^2$. As it can be noted from (44), the SNRs across all subcarriers are equal, which implies that this system has introduced frequency diversity to the OFDM system. However, if any of the subcarriers fades substantially, then $H$ for that subcarrier could be very close to zero. Consequently, the SNR for that subcarrier as well as all subcarriers will be close to zero. Therefore, this approach would be useful in mild frequency-selective channels. To remedy this problem several solutions can be adopted such as discarding the subcarriers with low $H$ values before the last FFT operation, add a fixed bias to the subcarriers in deep fade, or use the minimum mean squared error (MMSE) equalizer. Once the division by zero problem is solved, the system will experience a remarkable BER improvement over frequency-selective channels. The $\text{SNR}|_{\mathbf{H}}$ using other equalization techniques can be derived similar to the ZF approach. However, due to space limitation we present only the Monte Carlo simulation results.

## VI. NUMERICAL RESULTS

To demonstrate the security strength of the proposed approach, a general OFDM system is simulated over AWGN channel with a small $N \times N$ $\mathbf{P}$ matrix, which corresponds to the shortest block length of $L = 1$. The number of subcarriers $N = 256$, the data symbols are selected from a QAM constellations with 4, 16, and 64 levels, the SNR is set to 30 dB. The number of samples mixed $\mathcal{K}$ varies from 0 to 256. As depicted in Fig. 1, even if the attacker knows up 200 values of the correct order, it is still very difficult to get any reliable data as the SER is more than 90% for most QAM values. It is worth noting that the upper limit for the SER is equal to $(1-1/M)$, in such a case the receiver is just selecting any of the $M$ possible symbols randomly. The results of Fig. 1 show that this bound is almost achieved for $M > 4$ given that $\mathcal{K} \gtrsim 50$.

The BER performance in frequency-selective fading channels is depicted in Fig. 2. The adopted channel model corresponds to a moderate frequency-selective channel that consists of five multipath components with normalized delays of $[0, 1, 2, 6, 11]$ samples and gains of $[0.34, 0.28, 0.23, 0.11, 0.04]$. The mean-squared delay spread of the channel is equal to 6.37 and the multipath components' powers are modeled as independent Rayleigh random numbers. The channel parameters remain constant over one block of $N$ OFDM symbols, which corresponds to a static channel that is widely used to model powerline and broadcasting channels.

Fig. 2 presents the BER of the proposed and standard OFDM systems. The ZF results are obtained semi analytically using (44) and confirmed using Monte Carlo simulation, the MMSE BER results is obtained using simulation. As it can be noted from this figure, the improvement achieved using the ZF equalizer is very limited while a remarkable improvement of more than 15 dB is achieved using the proposed system with MMSE.

## VII. CONCLUSION AND FUTURE WORK

This work presented a novel physical layer security technique for OFDM based communication systems. The proposed system is based on using a permutation matrix of particular size to reorder the $\mathcal{K}$ OFDM samples in a random manner. Simulation results demonstrated that the system is highly immune against the chosen cipher-text attack as long as the permutation matrix changes pseudorandomly each transmission block. This requirement is not essential for other simpler attacks such as the chosen ciphertext attack. In addition to the security features of the proposed system, analytical and simulation results demonstrated that a significant BER improvement of more than 15 dB is achieved using the MMSE equalizer as compared to the standard OFDM.

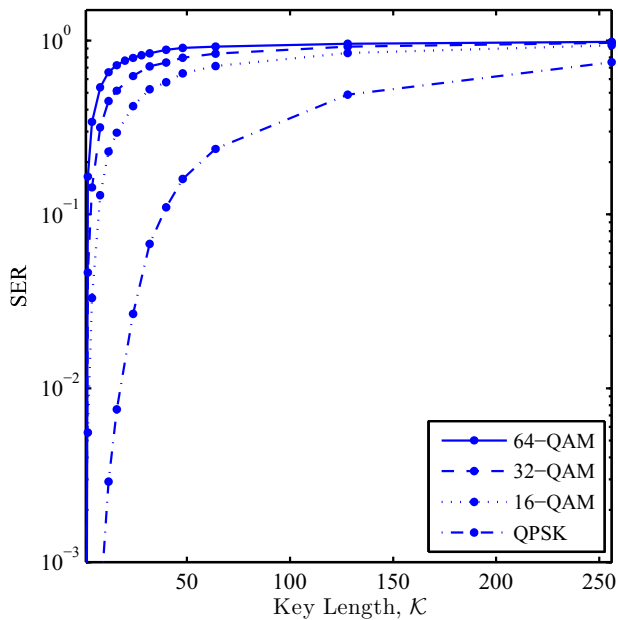

Fig. 1. Example-I, the SER using 64, 32, 16 and 4 QAM constellations for different values of $\mathcal{K}$, $L = 1$, out of order samples.

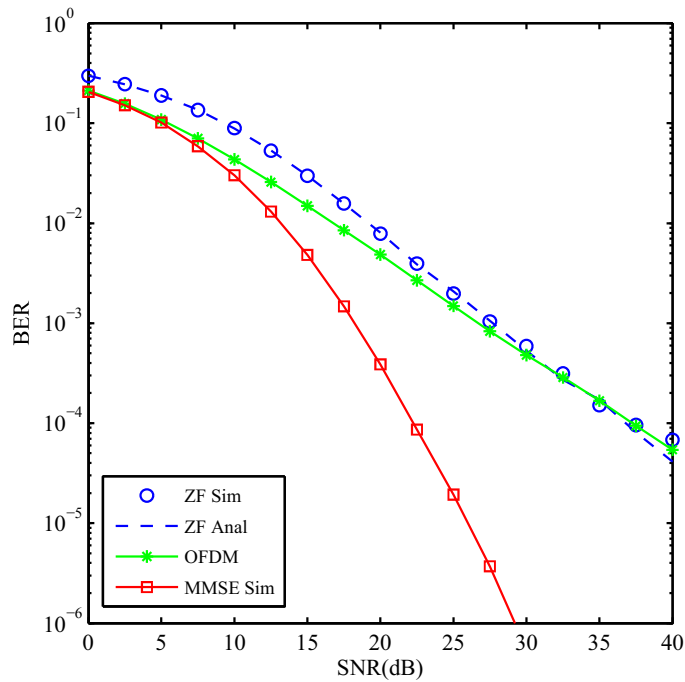

Fig. 2. The BER of the proposed system using ZF and MMSE equalizers in frequency-selective fading channels.

Our future work will focus on evaluating the performance of the proposed system using nonlinear functions and using different combining matrices. Moreover, the derivation of analytical or semi analytical BER using the MMSE and equalization methods will be considered as well.